\documentclass{astro-ph-1}

\usepackage{mathptmx}
\newcommand\ltsim{\,\lower0.7ex\hbox{$\stackrel{<}{\sim}$}\,}
\newcommand\gtsim{\,\lower0.7ex\hbox{$\stackrel{>}{\sim}$}\,}

\begin{document}

\fontfamily{ptm}
\title{Advances in Multi-Dimensional Simulation of \\ Core-Collapse Supernovae
\footnote{\uppercase{T}o appear in published proceedings of {\em
\uppercase{O}pen \uppercase{I}ssues in
\uppercase{C}ore-\uppercase{C}ollapse \uppercase{S}upernovae}, which
was conducted at \uppercase{T}he \uppercase{I}nstitute for
\uppercase{N}uclear  
\uppercase{T}heory, \uppercase{U}niversity of \uppercase{W}ashington, 
\uppercase{S}eattle, \uppercase{WA, USA,}
\uppercase{J}une, 2004.}
}

\author{F.~Douglas Swesty and Eric~S. Myra
\footnote{\uppercase{V}isiting \uppercase{F}ellows, 
\uppercase{T}he \uppercase{I}nstitute for \uppercase{N}uclear 
\uppercase{T}heory, \uppercase{U}niversity of \uppercase{W}ashington, 
\uppercase{J}une, 2004.}}

\address{Dept. of Physics \& Astronomy, \\
State University of New York at Stony Brook \\ 
Stony Brook, NY 11794--3800, USA\\ 
\begin{tabular}{ll}
E-mail:  & dswesty@mail.astro.sunysb.edu\\
&  emyra@mail.astro.sunysb.edu
\end{tabular}}

\maketitle
\fontfamily{ptm}
\abstracts{
We discuss recent advances in the radiative-hydrodynamic modeling of
core collapse supernovae in multi-dimensions.  A number of earlier
attempts at fully radiation-hydrodynamic models utilized either the
grey approximation to describe the neutrino distribution or utilized
more sophisticated multigroup transport methods restricted to radial
rays.  In both cases these models have also neglected the $O(v/c)$
terms that couple the radiation and matter strongly in the optically
thick regions of the collapsed core.  In this paper we present some
recent advances that resolve some shortcomings of earlier models.  }

\section{Introduction: The Supernova ``Problem'' }

For over a decade researchers have struggled to model the convective
post-bounce epoch of core collapse supernovae.  The
radiative-hydrodynamic flows that occur in the region below the
stalled prompt shock have held both promise and pitfall for the
supernova modeler.  The promise of this phenomenon is that it might
explain the long sought-after mechanism that converts the core bounce
into the observed explosion.  The pitfalls are legion, mostly
involving a complex convective flow structure that is
three-dimensional in nature and couples neutrinos to matter strongly.
For this reason there remain many open issues in modeling the
convective epoch of core collapse supernovae.

The supernova ``problem'' persists, despite more than four decades of
concentrated research.  The problem is this: We have no convincing
explanation as to how the core collapse, which ends the evolution of a
massive star, rebounds in such a way as to generate the explosion we
observe in nature.  The most realistic supernova models collapse and
rebound, but create a shock wave that doesn't eject matter, either on
the hydrodynamic timescale ($\sim 10$~ms), or the diffusive timescale
of the escaping neutrino radiation ($\sim$ 1--10~s), or on any other
timescale we can model.

This is not a problem of overall energetics.  The gravitational energy
released during core collapse and the subsequent neutron-star cooling
phase is several factors of $10^{53}$~erg.  In contrast, the kinetic
energy of the explosion required for consistency with observation is
only $10^{51}$~erg.  Instead of insufficient energy, the problem is
one of energy conversion and transport---how a sufficient portion of
the released gravitational energy is imparted to the material {\em
ejectus}, giving it the requisite kinetic energy.

It has been understood for many years that neutrinos play a vital role
in this process.  In fact, essentially the entire remaining 99\% of
released energy (that which is not converted to kinetic energy of the
matter) is radiated away as neutrinos.  Thus, an accurate treatment of
neutrino processes is a necessary component of any realistic model for
a supernova.

In this article, we present what we currently regard as the most
important issues in core-collapse supernova modeling.  In Sec.~2, we
discuss the major components that need to be part of any serious
modeling endeavor. In Sec.~3, we present an outline of V2D, our new
two-dimensional (2-D) supernova simulation code.  Section~4 contains
some preliminary results using this code.  Our conclusions are in
Sec.~5.

\section{The Components of a Supernova Simulation}

Broadly speaking, there are four main components to current supernova
simulation models: (1) hydrodynamics, to track the collapse, rebound,
and ejection of stellar material, (2) neutrino transport, to track the
production of neutrino radiation and to follow its propagation and
emission from the star, (3) nuclear microphysics, to describe the
diverse states of matter encountered throughout a simulation, and (4)
neutrino microphysics, to describe the reactions and interactions
involving neutrinos and matter.  It must be stressed that all of these
components are tightly coupled to one another.  Thus, the most
effective models are designed with this coupling built in {\it ab
initio}.
\vspace*{5mm}

{\bf Hydrodynamics.} For simplicity of implementation, it has been
customary for supernova codes to employ explicit hydrodynamics, with
either a Newtonian or a general relativistic formulation.  Implicit
algorithms have usually been avoided since they require the 
computationally expensive solution of large systems of non-linear
equations. 

Regardless of which approach is used, the hydrodynamic portion of the
problem requires solution of some form the following equations,
expressed here in Newtonian formalism:

\begin{equation} \label{eq:cont}
\frac{\partial \rho}{\partial t} +
{\pmb \nabla} \cdot \left( \rho {\bf v} \right) = 0
\end{equation}
\begin{equation} \label{eq:ye}
\frac{ \partial \left( \rho Y_e \right) }{\partial t} +
{\pmb \nabla} \cdot \left( \rho Y_e {\bf v} \right) = 
-m_b \sum_f \int d\epsilon 
\left( \frac{{\mathbb S}_{\epsilon}}{\epsilon} -
\frac{ \bar{{\mathbb S}}_{\epsilon}}{\epsilon} \right)
\end{equation}
\begin{equation} \label{eq:energy}
\frac{ \partial E}{\partial t} +
{\pmb \nabla} \cdot \left( E {\bf v} \right) +
P {\pmb \nabla} \cdot {\bf v} =
- \sum_f \int d\epsilon 
\left( {\mathbb S}_{\epsilon} + \bar{{\mathbb S}}_{\epsilon} \right)
\end{equation}
\begin{equation} \label{eq:mom}
\frac{ \partial \left( \rho {\bf v} \right) }{\partial t} +
{\pmb \nabla} \cdot \left( \rho {\bf v}{\bf v} \right) +
{\pmb \nabla} P + \rho {\pmb \nabla} \Phi +
{\pmb \nabla} \cdot 
\left\{ \sum_f \int d\epsilon \left( {\mathsf P}_{\epsilon}
+ \bar{{\mathsf P}}_{\epsilon} \right) \right\} 
= 0.
\end{equation}

Equation (\ref{eq:cont}) is the continuity equation for mass, where
$\rho$ is the mass density and ${\bf v}$ is the matter velocity, and
where these quantities, and those in the following equations, are
understood to be functions of position {\bf x} and time $t$. Equation
(\ref{eq:ye}) expresses the evolution of electric charge, where $Y_e$
is the ratio of the net number electrons over positrons to the total
number of baryons.  In the presence of weak interactions, the right
hand side is non-zero to account for reactions where the number of
electrons can change.  Here, we express the net emissivity of a
neutrino flavor (of energy $\epsilon$) and its antineutrino by
${\mathbb S}_{\epsilon}$ and $\bar{{\mathbb S}}_{\epsilon}$,
respectively.  This expression is integrated over all neutrino
energies and summed over all neutrino flavors $f$.  The mean baryonic
mass is given by $m_b$.  Evolution of the internal energy of the
matter is given by the gas-energy equation, Eq.~(\ref{eq:energy}),
where $E$ is the matter internal energy density and $P$ is the matter
pressure.  Again, the right hand side of this equation is non-zero
whenever energy is transferred between matter and neutrino radiation
as a result of weak interactions.  We note that it is also possible to
substitute for Eq.\ (\ref{eq:energy}) an expression for the
evolution of the {\em total} matter energy (internal plus kinetic plus
potential).  Finally, Eq.\ (\ref{eq:mom}) expresses gas-momentum
conservation, where $\Phi$ is the gravitational potential, and
${\mathsf P}_{\epsilon}$ and $\bar{{\mathsf P}}_{\epsilon}$ are
radiation-pressure tensors for each energy and flavor of neutrino and
its anti-neutrino, respectively.

These equations must be discretized for solution within a
computational framework.  Traditionally, with one-dimensional models,
it has been convenient to use Lagrangean methods, in which a
computational mesh strictly co-moves with the mass elements of the
fluid. With the advent of multi-dimensional models, however, it is
common to use Eulerian hydrodynamics, where the mesh is fixed in an
inertial frame of reference.  This is because purely Lagrangean
methods are difficult to implement in multi-dimensional schemes
without the mesh suffering distortion and entanglement in convectively
active regions.  

For all the benefits of Eulerian meshes, they also present a number of
thorny issues. This is especially true for spherical polar meshes, the
most natural choice for supernova modeling.  The most obvious issue is
the coordinate singularity that exists when the polar angle, $\theta
\rightarrow 0$. In addition, polar meshes exacerbate the problem of the
timestep-restricting Courant-Friedrichs-Levy (CFL) condition at the
center of the core.  To deal with these issues, there are numerous
resolutions and combinations of resolutions under active
consideration. These include implicit methods, unstructured meshes,
body-fitted meshes, and adaptive mesh refinement (AMR).

As mentioned above, it is also necessary to choose between a total
energy and an internal energy formulation .  For the supernova
problem, an internal energy formulation, as given in Eq.\
(\ref{eq:energy}), is preferred.  This is because much of the energy is
internal, as opposed to kinetic.  Solving the gas-energy equation
helps insure an accurate calculation of the entropy, which is
critical in degenerate regimes where a small change in energy can lead
to a large change in temperature.

The hydrodynamic algorithm must also have convergence properties that
can deal with a realistic equation of state.  This is particularly
important in the regions of non-convex phase changes, such as the
transition between nuclei and continuous nuclear matter.  
\vspace*{5mm}

{\bf Neutrino Transport.} This component is the most difficult to
implement in a supernova model and the most time-consuming
computationally. This is because supernova neutrinos cannot, in
general, be described by an equilibrium distribution function.  A
solution requires a complete phase-space description of each
neutrino's position and momentum.  To obtain such a solution, one must
solve the six-dimensional Boltzmann Transport Equation or some
reasonable approximation thereof.  This extra dimensionality easily
leads to the transport calculation completely dominating a simulation
in terms of computer memory, execution time, and I/O requirements.

The Boltzmann Transport Equation (BTE) can be expressed in terms of
the radiation intensity, $I = I(\epsilon, {\bf x}, {\pmb \Omega}, t),$
where $\epsilon$ is the energy of a neutrino, ${\bf x}$ its position,
and ${\pmb \Omega}$ the solid angle into which the neutrino radiation
is directed. In terms of $I$, the Newtonian BTE can be expressed as
\begin{equation}\label{eq:bte}
\frac{1}{c} \frac{\partial I}{\partial t} +
{\pmb \Omega} \cdot {\pmb \nabla} I + \sum_i
a_{i} \frac{\partial I}{\partial p_{i}} =
\left( \frac{\partial f}{\partial t} \right)_{{\rm coll.}},
\end{equation}
where $a_{i}$ is the $i^{th}$ component of the matter acceleration and
$p_{i}$ the $i^{th}$ component of the momentum of the neutrino.  The
right hand side of Eq.\ (\ref{eq:bte}) lumps together the
contributions from all interactions that a neutrino might experience
and is collectively referred to as the collision integral.

A storm of issues faces one who implements a neutrino transport
algorithm.  Mezzacappa and Bruenn\cite{mb93a} have the only ``full''
solution to the BTE implemented in supernova simulations and then only
with one-dimensional hydrodynamics.  Upon moving to multi-dimensional
models, the full solution of the BTE becomes yet more challenging to
implement and more time-consuming to compute. However, this is the way
that the field must ultimately go. (Livne and colleagues
purport\cite{livne} to implement a two-dimensional $S_n$ solution of the
BTE.  However, this solution omits critical matter-radiation coupling
terms and no numerical details of the method have been disclosed.)

In the meantime, a number of approximate transport moment methods have
emerged.  The most successful of these is use of a finite series of
angular moments of the BTE.  When this approach is taken, a limiting
scheme is then required to close the resulting equations.
Breunn\cite{bruenn85} implemented a $P_1$ scheme, with flux-limiting.
Myra~{\em et al.}\cite{MBHLSV} closed the zeroth angular moment of the
BTE by implementing the Levermore and Pomraning flux limiter\cite{LP}.
Bowers and Wilson\cite{BW82} also used a flux-limiting scheme of their
own device. More recently, Rampp and Janka\cite{RJ} have implemented a
variable-Eddington-factor approach to solve the first two angular
moments of the BTE.  In all the above cases, however, the
implementation has been made in only one spatial dimension.
(Janka~{\em et al.}\cite{janka03} are developing a two-dimensional
Boltzmann transport code (MuDBaTH), but the numerical details are as
yet unpublished.)

Since all the schemes noted so far derive monochromatic transport
equations, yielding a separate equation for each neutrino energy, they
are referred to as multi-group schemes.  Those that combine
multi-group and flux-limiting are known as multi-group flux-limited
diffusion (MGFLD) schemes.

A much simpler alternative to multi-group schemes is so-call ``grey''
transport, which is derived by integrating the BTE over both neutrino
energy and angle.  To perform these integrals, one must assume a
spectral shape for the neutrino distribution.  Typically, this
requires defining an arbitrary neutrino ``temperature,'' and assuming
that neutrinos can be parameterized by some kind of equilibrium
Fermi-Dirac distribution.  Among relatively recent models, this was
first implemented by Cooperstein, van den Horn, and Baron,\cite{cvdhb}
and later by Swesty\cite{fdsgrey} and by Herant~{\em et al.}\cite{engine}
This approximation is still in active use by the latter group.\cite{fw}

Grey schemes have numerous shortcomings.  First, work with multi-group
schemes has shown that in areas where accurate neutrino transport
is critical, neutrinos do not assume any kind of distribution that can
parameterized once and for all as required by grey transport.
Spectral distributions constantly evolve and, thus, a multi-group
description is required to obtain even a qualitatively correct
description. More troubling, Swesty\cite{fdsgrey} has shown that by
adjusting the grey parameterization within very small bounds, it is
possible to ``dial'' an explosion (or failed explosion) with the
appropriate choices of these unknowable and unphysical tuning
parameters.  Hence, although grey codes have utility for making a
sweeping exploration of parameter space, any scientific conclusions
that rely on them should be viewed as highly suspicious, and not
regarded as in any way definitive.

Regardless of which transport scheme is implemented, another critical
issue that must be faced is matter-radiation coupling.  Coupling
occurs in Eqs.~(\ref{eq:ye})--(\ref{eq:mom}) for lepton number, energy,
and momentum evolution.  Coupling that occurs on the right-hand
side of these equations has a conceptually simple analytic structure.
However, momentum transfer is more troublesome in approximate schemes.
This is because the radiation momentum equation is often truncated in
a way that makes the accuracy of the calculated momentum transfer less
certain.

Matter-radiation coupling also enters implicitly in the neutrino
transport equation through spectral rearrangement terms and in the
dynamic diffusion term.  Both these terms are frequently and
erroneously neglected in supernova models, even though the dynamic
diffusion term is the leading order contribution in optically-thick
regions.
\vspace*{5mm}

{\bf Equation of State.} Adequate modeling of stellar-core collapse
requires an equation of state (EOS) that handles a density range of
roughly $10^5$--$10^{15}$~g ${\rm cm^{-3}}$, a temperature range of
0.1--25~MeV, and an electron-fraction range of 0.0--0.5.  The EOS
must also be able to handle different regimes of equilibrium states.
Throughout most of the core, the material is in nuclear statistical
equilibrium (NSE) and is usually modeled by one of the NSE equations
of state. Although the gross features of nuclear matter are thought to
be well-understood, there is still much open ground for investigation.
Fertile regimes for such work include the EOS at supernuclear
densities.  In addition, little is known about the nature of nuclei at
subnuclear densities when $Y_e$ is small.

Matter in the silcon shell and beyond does not attain NSE until the
bounce-shock wave passes through it.  Dealing with the transition
between NSE and non-NSE EOS's and with the network of nuclear
reactions that joins them is a challenge that is only beginning to be
addressed.\cite{hix}
\vspace*{5mm}

{\bf Neutrino Microphysics.} Since the energetics of a core-collapse
supernova is primarily a neutrino phenomenon, it is necessary to have
correct opacities and rates for the various neutrino processes that
are important.  The collection of reactions that are important, or
possibly important, to the supernova problem is rich and has evolved
through the years. Arguably the most important development came with
the discovery of weak neutral currents, from which it could be
inferred that the dominant contribution to neutrino opacity in a
collapsing stellar core is from coherent elastic scattering of
neutrinos from nuclei.\cite{freed}

The list of possible neutrino interactions is nearly endless, but
those of demonstrated importance include the coherent scattering just
mentioned, as well as conservative scattering from free nucleons.
Also of undisputed importance are electron capture by protons (and
protons bound in nuclei), neutrino production through
electron-positron pair annihilation, and neutrino-electron scattering.

In recent years, with the experimental evidence pointing strongly to
the existence of neutrino oscillations, it is also important to
investigate the possible role of flavor-changing interactions to the
supernova problem.  Investigation into this has begun,\cite{sb} but
has yet to be incorporated in a detailed simulation.

\section{V2D: A New Code for Two-Dimensional Radiation Hydrodynamics}
Our new radiation-hydrodynamic simulation code, V2D, is a
two-dimensional, Newtonian, pure Eulerian, staggered-mesh code based
on a modified version of the algorithm for ZEUS-2D by Stone and
Norman.\cite{sn1,sn2,sn3} Following Stone and Norman, it has been
designed for use in a general orthogonal two-dimensional geometry,
which makes its utility extend beyond the supernova problem.

V2D is an entirely new implementation, coded according to the
Fortran~95 standard.  It is a distributed-memory parallel code that
uses calls to MPI-1 for message passing between processes. It has been
designed for easy portability between computing platforms and
currently runs on systems ranging from as small as a Linux-based laptop
to as many as 2048 processors of an IBM SP.  To aid in this
portability, the input and output is formatted using parallel HDF5,
which is built on the MPI-I/O portion of the MPI-2 standard.

One of the major design goals of V2D is componentization and, to
adhere to this principle, we insist on completely separating
microphysics from the numerical implementation of our
radiation-hydrodynamics algorithm. This isolation of mathematics and
computational science from physics has allowed significant
contributions from applied mathematicians to enhancing the performance
of our code.

The V2D algorithm relies on operator splitting, with advection steps
split from source-term steps.  Hydrodynamic and neutrino-transport
source-term steps and coupling are interleaved.  At the start of each
simulation timestep, the gravitational mass interior to each point is
calculated.  Since the collapsed core is nearly spherically symmetric
and highly condensed, we approximate the gravitational mass assuming
that mass interior to the point of interest is in a spherically
symmetric distribution. In future versions of our code, we will
implement a more accurate Poisson solver to calculate the (slightly)
non-spherical gravitational potential.

In a break from Stone and Norman's method, V2D next performs the
advection sweeps in the radiation-hydrodynamic quantities (mass
density, matter internal energy, velocities and momenta, electron
fraction, and neutrino distributions).  Following this, a neutrino
transport step is performed for each flavor and each matter-radiation
energy exchange is calculated.

The matter pressure is next updated, upon which the gravitational and
matter- and radiation-pressure forces are applied to the matter.
Artificial viscosity is calculated next and its contributions applied
to the fluid.  Finally, the gas-energy equation is solved.

This procedure is repeated for each timestep in a simulation, with the
provision that advection sweeps are ordered alternately according to
timestep ({\it i.e.,} $x_1$-direction first, followed by $x_2$, or
{\it vice versa}).

\subsection{Neutrino Transport Implementation}
As an extension of earlier work by us,\cite{MBHLSV,SSS} we
implement neutrino transport by taking the zeroth angular moment of
the BTE to yield the following neutrino monochromatic energy equation
in the co-moving frame:
\begin{equation}\label{eq:bte0}
\frac{\partial E_{\epsilon}}{\partial t} +
{\pmb \nabla} \cdot \left( E_{\epsilon} {\bf v} \right) +
{\pmb \nabla} \cdot {\bf F}_{\epsilon} -
\epsilon \frac{\partial}{\partial \epsilon} 
\left( {\mathsf P}_{\epsilon}:
{\pmb \nabla} {\bf v} \right) = {\mathbb S}_{\epsilon},
\end{equation}
where $E_{\epsilon}$ is the neutrino energy density per unit energy
interval at position ${\bf x}$ and time $t$, ${\bf F}_{\epsilon}$ is
the neutrino energy flux per unit energy interval, and ${\mathsf
P}_{\epsilon}$ and ${\mathbb S}_{\epsilon}$ are as defined
earlier. The expression ${\mathsf P}_{\epsilon}:{\pmb \nabla}
{\bf v}$ indicates contraction in both indices of the second-rank
tensors ${\mathsf P}_{\epsilon}$ and ${\pmb \nabla} {\bf v}$.
There is a corresponding equation to describe the antineutrinos. This
pair of equations is repeated for each neutrino energy $\epsilon,$
and neutrino flavor.  We currently track electronic, muonic, and
tauonic neutrinos.

Equation (\ref{eq:bte0}) is closed using Levermore and Pomraning's
prescription for flux-limited diffusion,\cite{LP} which allows us to
express ${\bf F}_{\epsilon}$ as
\begin{equation}
{\bf F}_{\epsilon} = -D_{\epsilon} {\pmb \nabla}  E_{\epsilon},
\end{equation}
where $D_{\epsilon}$ is a ``variable'' diffusion coefficient that
varies in such a way as to yield the correct fluxes for the diffusion
and free streaming limits and an approximate solution in the
intermediate regime.  This prescription also provides the elements of
the radiation-pressure tensor ${\mathsf P}_{\epsilon}$.

Presently, we employ the same prescriptions for electron capture and
conservative scattering that we have used in the
past.\cite{MBHLSV,bruenn85} The rates for these process are
calculated on the fly within the course of a simulation.  Our model
also implements neutrino production via electron-positron pair
annihilation, as in Yueh and Buchler\cite{yb76} and
Bruenn\cite{bruenn85}. Neutrino-electron scattering has been
implemented, but is not currently turned on in the preliminary results
we present here.  These latter two sets of processes use tables of
precomputed rates, which are interpolated via a tri-linear
interpolation scheme over neutrino energy $\epsilon$, temperature 
$T$, and electron chemical potential $\mu_e$.

Since the neutrino CFL restriction on a transport timestep is far too
restrictive to permit an explicit solution, we use a purely implicit
method to solve the transport. The equations comprising the
description of each neutrino-antineutrino species are assembled in
matrix form. We note that the second (advective) term in
Eq. (\ref{eq:bte0}) is omitted from this process since it has been
already treated during the operator splitting of the advective step
described above. Blocking terms arising from Fermi-Dirac statistical
restrictions on final neutrino states make this a system of non-linear
equations.  Fortunately, the system is sparse, which makes it amenable
to solution by sparse iterative methods. A nested procedure is used,
employing Newton-Krylov methods\cite{SSS}.  In the innermost loop, a
linearized system is solved using preconditioned Krylov-subspace
methods.  The outer loop uses a Newton-Raphson iterative scheme to
resolve the non-linearity of the system.  Besides being an effective
general procedure for sparse systems, our implementation of parallel
preconditioners also insures that is amenable to large-scale solution
on parallel architectures.  This is the chief reason our code
exhibits its high degree of scalability across many platforms.

\subsection{Equation of State}
V2D is designed to use an arbitrary equation of state and we use
several in the course of testing the code.  For production runs,
however, we use the Lattimer-Swesty EOS\cite{LS,LLPR} in
tabular form.  The thermodynamic quantities are tabulated in terms of
independent variables, density, $\rho$, temperature, $T$, and electron
fraction, $Y_e$.  We have tabulated this EOS in a thermodynamically
consistent way according to the prescription in Swesty\cite{FDSTCT}.
(We refer to this combination collectively as LS-TCT.)  We note that
although the Lattimer-Swesty EOS is commonly used, and tabulations of
it are also common, most tabulations are not constructed in such a way
as to {\em guarantee} thermodynamic consistency.  When
non-thermodynamically-consistent tables are used, spurious entropy can
be generated or lost. Such problems have been sometimes incorrectly
attributed to the Lattimer-Swesty EOS, rather than erroneous
tabulation of the otherwise consistent EOS.

To guarantee tabular thermodynamic consistency, LS-TCT uses bi-quintic
Hermite interpolation in the Helmholtz free energy $F$, as a function
of $T$ and $\rho$.  Functional dependence on $Y_e$ changes slowly
enough to permit linear interpolation.  This procedure is required
since satisfaction of the Maxwell relations requires consistency among
the second derivatives of $F$.  In addition, we desire fidelity of the
interpolation and continuity of derivatives to the underlying tabular
data for both $F$ and its derivatives, $(\partial F/\partial
T)_{\rho}$ and $(\partial F/\partial \rho)_T$.

Apart from thermodynamic considerations, we also want to insure that
there are no discontinuities that might cause difficulties in the
hydrodynamics. Hence, LS-TCT also insists that interpolations of the
second derivatives of $F$ approach the correct tabulated values.
(This is the equivalent of saying that we require the derivatives of
pressure and internal energy with respect to temperature and density
be continuous.)

A final requirement is that we wish the interpolation function and its
first and second derivatives be continuous across table-cell
boundaries.  This insures that nothing untoward happens as a fluid
element migrates from one thermodynamic regime of interest to another.

\subsection{Software Validation and Verification}

When engaged in a major project, such as V2D, it is important for
software developers to be constantly vigilant about the quality of
software being produced.  It is important that the software meet the
requirements that it is intended to address (validation) and that the
code yields correct answers (verification).

We take these issues seriously and, in an effort to address them, have
implemented strict source-code control and testing procedures to
ensure that our software meets our rigorous standards.  One of the
most important elements of our program has been the implementation of
a suite of regression tests, which we currently run four times daily.
This is not a static suite, but is constantly growing.  Our eventual
aim is to cover every major element of code in V2D.  Currently our
suite consists of about two dozen separate problems that include tests
of the hydrodynamics, neutrino transport, parallel solvers, message
passing, and parallel I/O.  Wherever possible, we try to include
problems with analytic or at least verifiable solutions.

Although implementation of these procedures is labor intensive and
time-consuming, we feel it is a justifiable investment.  With current
regression tests, our procedures have already been effective in
finding errors in our code.  They have also served as an effective
safeguard against introducing new errors as we continually enhance
V2D's functionality.

\section{Initial Results}

We are using V2D to carry out our first 2-D multigroup models of the
post-bounce epoch.  To date, simulations have not reached a sufficient
time that would allow us to be certain about whether an explosion is
obtained.  Nevertheless, the results warrant some discussion as they
reveal important features of the post-bounce epoch.

\subsection{Initial Model}

For a progenitor model we employ the widely-used Woosley and
Weaver\cite{ww} S15S7B2 $15 M_{\odot}$ progenitor.  Much previous work
has focused on the evolution of this progenitor through collapse, core
bounce, and convective phases.  The Fe core, the Si shell, and a
portion of the O shell area are zoned into a 256 radial-mass-zone mesh
with zoning that is tuned (by trial and error) so as to yield a high
spatial resolution grid in the proto-neutron star and the inner 200 km
of the collapsed core at bounce.  This tuned zoning sets up a radial
grid that is compatible with subsequent 2-D Eulerian simulations.  The
neutrino-energy spectrum, ranging from 0--375 MeV, is discretized into
20 energy groups with group widths that increase geometrically with
energy so as to resolve accurately the Fermi surface of the electrons
and neutrinos in the proto-neutron star.  The initial values for $T$,
$\rho$, and $Y_e$, are interpolated from the original S15S7B2 data
onto the Langragean mass grid and the initial radial coordinates of
each mass shell are computed consistently with density.  The neutrino
energy densities $E_\nu$ are initialized to a small non-zero value
that yield an initial neutrino luminosity that is many orders of
magnitude below what the precollapse thermal pair-production
luminosity would be.  When the simulation is started in its
pre-collapse quasi-static phase, the luminosity stabilizes within a
light crossing time (5-10 ms).

\subsection{Lagrangean Collapse Calculations}

The initial model is collapsed using a 1-D Newtonian Lagrangean
radiation hydrodynamics code RH1D that uses the mass and energy meshes
described above.  The evolution algorithm for each timestep utilizes
operator splitting to first carry out a Lagrangean hydrodynamics step
followed by Lagrangean neutrino evolution steps for each of the three
neutrino flavors.  After each Lgrangean neutrino evolution step the
matter internal energy and electron fraction are corrected for any
energy and lepton number exchange that has occurred.

The model contains neutrino microphysics as described in
Bruenn\cite{bruenn85} with two exceptions.  The neutrino-nuclei
scattering opacity has been modified to take into account the
form-factor introduced by Burrows, Mazurek, and Lattimer\cite{bml}
and we have neglected the effect of neutrino/anti-neutrino
annihilation.  Full neutrino-electron scattering as described in this
paper is included in the code but is not turned on in the model
described in this paper.  Additional effects such as nucleon recoil,
ion-ion correlations, etc.  are being considered as follow-ons to this
baseline model.

The EOS utilized is the TCT tabularized version of the nuclear EOS
Lattimer-Swesty (LS-TCT) with the $k=180$~MeV parameter set, with the
exception of the electron EOS that has been updated to improve
accuracy and the range of applicability.  The original LS EOS chose a
value for the alpha particle binding energy $B_\alpha = 28.3$~MeV that
did not correctly account for the neutron-proton mass difference.
This parameter has been subsequently corrected but in the model
presented in this paper, the original value has been retained so as to
allow comparison to other work.

Using RH1D with the microphysics and meshes described above the core
collapses in approximately 233 ms.  The central conditions at bounce
are approximately $T\approx 10.3$~MeV, $\rho \approx 2.71\times
10^{14}~{\rm g~cm^{-3}}$, $Y_e \approx 0.306$, and $Y_\ell \approx
0.37$.  These conditions are in good agreement with those obtained in
Lagrangean MGFLD and MGBT models
\cite{bruenn85,MBHLSV,slm93,mb93b}.  

\subsection{Eulerian 2-D Calculations}

\begin{figure}[ht]
\epsfxsize=10cm   
\centerline{\epsfxsize=3.5in\epsfbox{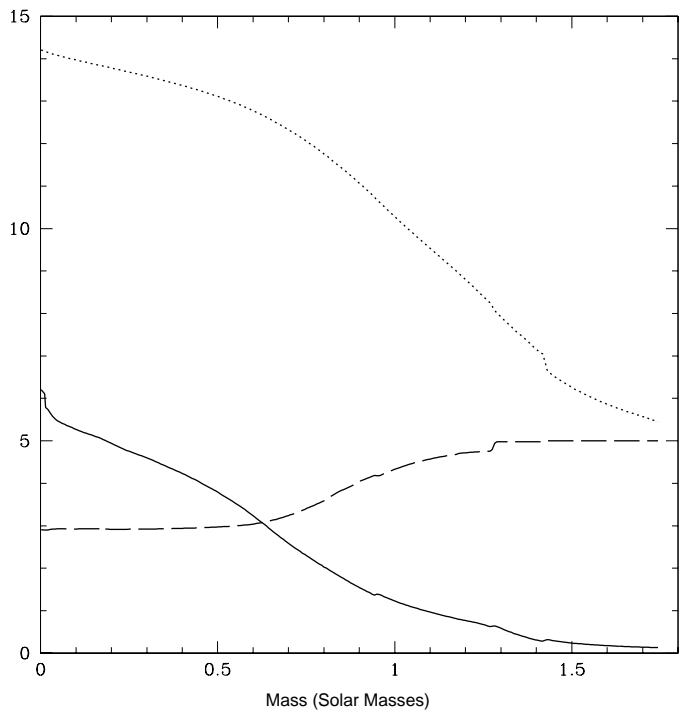}}
\caption{Initial radial profile for 2-D simulation. Solid line indicates
temperature in units of MeV.  Dashed line indicates $10\times Y_e$.
Dotted line indicates $\log_{10}(\rho)$ in units of g ${\rm cm^{-3}}$.}
\label{fig:fig1}
\end{figure}

Our 2-D models are carried out in spherical polar coordinates.  The
initial conditions for our 2-D simulations are taken from the 1-D
Langrangean simulations as the central density of the core reaches the
nuclear saturation density.  The $T$, $\rho$, and $Y_e$ profiles at
this point are shown in Figure \ref{fig:fig1} and are taken
at approximate time of $233.087$ ms.
The choice of this epoch to begin our 2-D models was made for two
reasons.  First, the prompt shock propagates and stalls into a
quasi-static equilibrium on the Eulerian mesh.  We have found that if
we attempt the transition from a Lagrangean algorithm to an Eulerian
algorithm at the point where the shock has stalled, there will be
noticeable differences in the quasi-static equilibrium caused by the
presence of the nuclear ``pasta'' phase transition.  These differences
in the equilibrium point can set off spurious unphysical shocks.  By
choosing to let the prompt shock propagate a stall on an Eulerian
mesh, we avoid this problem.  The second reason for choosing the
initial point for our 2-D models near the moment of bounce is that the
radial coordinates of the zones in the outer part have achieved
desirable values for an Eulerian simulation.

The radial zoning for the 2-D models is taken directly from the radial
coordinates of the 1-D Langrangean model zones.  In this way we avoid
any need for remapping of data between radial zones.  This eliminates
the introduction of any spurious forces into the initial conditions
for the 2-D models.  The initial data for $T$, $\rho$, $Y_e$, $v_r$,
and the neutrino energy densities $E_\nu$ and $\bar{E}_\nu$ for each
neutrino flavor are taken directly from the 1-D Langrangean model.
The initial velocity in the $\theta$ direction is set to zero.  The
data are mapped in the polar-angular direction in a spherically
symmetric fashion.  The angular grid consists of 256 zones uniformly
spaced in angle over the range of $0 \le \theta \le \pi$.  We place a
small sinusoidal perturbation in the electron fraction $Y_e$ to seed
convection in the region between 100 and 200 km of the form
$(Y_e)_{{\rm perturb}} = (Y_e)_{{\rm Lagrangean}} + C_p \sin(4\theta)$
where $C_p = 10^{-6}$.  The energy group structure is also left
unchanged from the 1-D Langrangean runs so there is no need to remap
the data in the energy dimension.  In the remainder of this paper we
shall refer to this model as Production Run 37 (PR37).

The use of spherical polar coordinates in combination with explicit
hydrodynamic algorithms to model spatial domains that include the
origin gives rise to a numerical stability problem.  At $r=0$ the
spherical coordinate system is degenerate and all zones that include
the origin as a vertex should be in instantaneous sonic communication
with one another.  However, standard explicit numerical
finite-difference, finite-volume, or finite-element techniques are
limited to nearest-neighbor type spatial coupling and do not include
numerical coupling between all zones containing a given vertex.  In
order to circumvent this problem we numerically introduce ``baffles''
into the center of the collapsed core, as though it were a tank of
fluid, to prevent movement of fluid in the angular direction inside a
certain radius.  Since no fluid movement occurs in the $\theta$
direction inside the baffle radius, there is no CFL restriction based
on the zone size in the $\theta$ direction for zones inside that
radius.  Nevertheless, the zones on either side of the baffle are
sonically connected as sound waves flow around the outer edge of each
baffle.  We strive to keep the baffle radius small, so that the flow
in the $\theta$ direction remains unimpeded in any region where
convective instabilities may develop. For the mode described in this
paper the baffle radius is approximately 8.5 km which yields an
average CFL timestep of about $5\times 10^{-7}$~s.  As we will see,
this baffle radius is well inside of any proto-neutron star (PNS)
instability region.

The 2-D calculations have been carried out on the IBM-SP system at the
National Energy Research Scientific Computing Center (NERSC).  The
models are run on 1024 processors with parallelism handled via message
passing via calls to MPI libraries.  Model PR37 required approximately
50,000 processor-hours of CPU time to reach a simulation time of 16 ms
post-bounce.

\subsection{The Onset of Convection}

The 2-D models are carried out from the point of bounce.  As expected
the prompt shock weakens while propagating outward and finally stalls.
In the 2-D models, the radius of the shock at 5 ms post-bounce is near
60 km and propagating outwards very slowly.  This is in good agreement
with our 1-D Lagrangean models.

\begin{figure}[ht]
\epsfxsize=10cm   
\centerline{\epsfxsize=4.1in\epsfbox{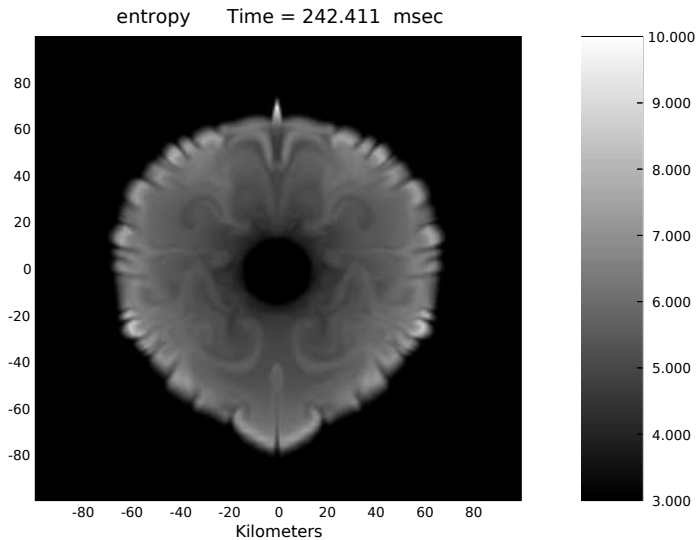}}
\caption{The entropy per baryon for model PR37 at approximately 
9 ms after core bounce.}
\label{fig:fig2}
\end{figure}

One difference that we see from previous works including the grey
models carried out by Swesty\cite{fdsgrey} is that convective
instabilities are born much earlier when the 2-D models are
initialized near the point of core bounce.  By 10 ms after bounce, the
model has developed two separate unstable layers.  The outer layer,
shown in Fig.~(\ref{fig:fig2}), is the classic entropy driven
Rayleigh-Taylor convection.

\begin{figure}[ht]
\epsfxsize=10cm   
\centerline{\epsfxsize=2.8in\epsfbox{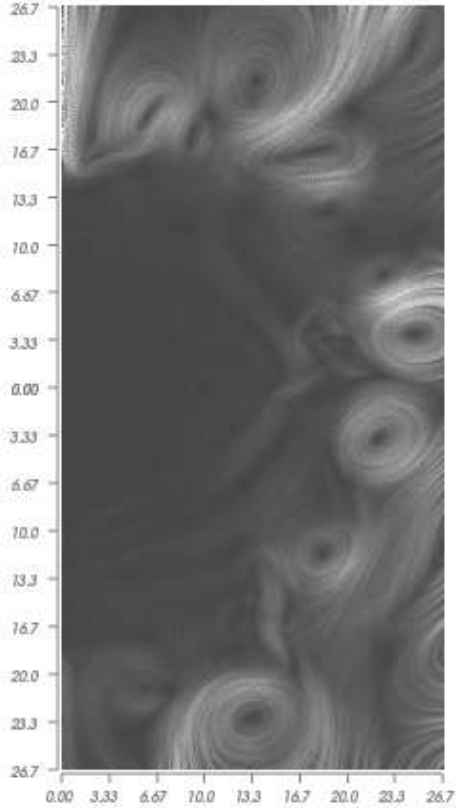}}
\caption{The velocity structure of the PNS instability layer
in model PR37 at t=242.411 ms.  The
image depicts the instantaneous structure of the
velocity field by means of a texture map
visualization technique known as
Lagrangean-Eulerian Advection (LEA).}
\label{fig:fig3}
\end{figure}

The inner layer, shown in Fig.~(\ref{fig:fig3}), 
seems to be an instability in the outer layers of the proto-neutron
star similar to those seen in 2-D simulations carried out by by Keil
{\em et al.}\cite{keil96} and Mezzacappa {\em et al.}\cite{mezz98}.  


There is controversy about the existence of PNS instabilities.
Originally, the work by Wilson and Mayle\cite{wm88,wm93} claimed to
find doubly-diffusive instabilities in the region below the
neutrinosphere.  Later work by Breunn and
collaborators\cite{bm94,bmd95,bd96} cast doubt on the existence of
such phenomena.  The simulations of Keil,\cite{keil96} which utilized
the grey flux-limited diffusion approximation to transport neutrinos
along radial rays, found a PNS instability that grew over the time
period of approximately one second to encompass the entire
proto-neutron star.  In contrast, the subsequent work of Mezzacappa
{\em et al.}\cite{mezz98} found a PNS instability that quickly damped
out within a short time.  The Mezzacappa {\em et al.} simulation
utilized a multigroup flux-limited $P_1$ approximation to transport
neutrinos along radial rays outward from an inner radius of r=20 km.
The inner boundary conditions in this simulation were established as
time-dependent data from the 1-D Langrangean code of
Bruenn\cite{bruenn85}.  It is important to note that neither the
simulations of Keil {\em et al.}\cite{keil96} or Mezzacappa {\em et
al.}\cite{mezz98} took into account the fully radiation-hydrodynamic
coupling via the compression and dynamic diffusion terms that are
present in Eq. (\ref{eq:bte0}).  Our simulations have confirmed the
expected result that the effects of the dynamic diffusion term
dominate the radiative diffusion term in the regions in which the
optical depth is large.

Figure 
\ref{fig:fig3} shows the instantaneous structure 
of the velocity field in model PR37 at the same
time as the data shown in Fig.~(\ref{fig:fig2}).
The velocity field is illustrated by means of a
Lagrangean-Eulerian Advection (LEA) visualization
technique that shows the direction of the vectors
as streaks.  One can clearly see eddies associated
with the PNS instability layer at a radius of
about 20--25 km.  This is well outside the 
baffle radius of 8.5 km.  In fact, there are approximately
a minimum of 40 radial zones separating the innermost of the
vortices and the outer edge of the baffles.  We do not believe
that the baffles in any way impede the dynamics of the vortices.
Nevertheless other simulations are underway where the baffle radius
is made substantially smaller to verify this claim.  We are also
developing an implicit hydrodynamic algorithm that will avoid the 
need for baffles altogether. 

Whether the PNS instability will grow or diminish with time is as yet
unclear since we have only evolved model PR37 to a time of
approximately 16 ms at the time of this writing.
\begin{figure}[ht]
\epsfxsize=10cm
\centerline{\epsfxsize=2.8in\epsfbox{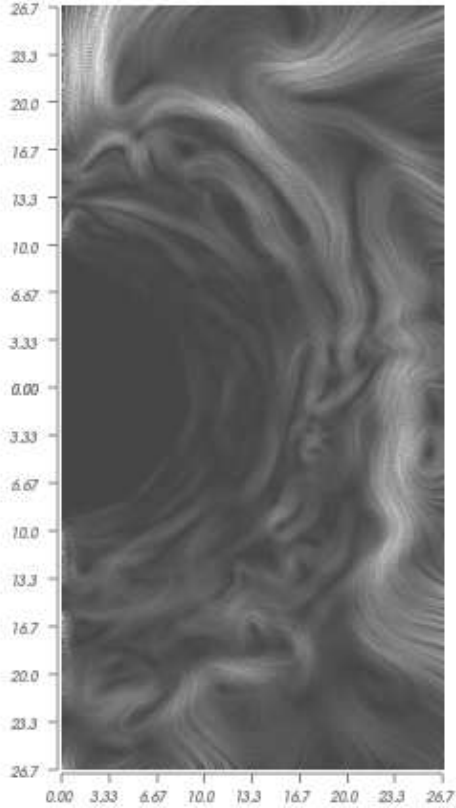}}
\caption{The velocity structure of the PNS 
instability layer in model PR37 at t=249.177 ms.}
\label{fig:fig4}
\end{figure}
The velocity structure at that time is shown in Fig.~(\ref{fig:fig4}),
which clearly reveals that much coherency has been lost in the
vorticial structure of the PNS instability layer.  This seems
indicative of the decay of this PNS instability in a fashion similar
to that described by Mezzacappa {\em et al.}\cite{mezz98}.  However,
it is necessary to evolve this simulation substantially farther
in time before any definitive statements can be made about the
long-term behavior of this sector of the proto-neutron star.  During
this relatively short timescale the outer convective zone seen in
Fig.~(\ref{fig:fig2}) does not exhibit significant growth.  With
evolution to the next 30 ms, we should be able to make
comparative statements regarding model PR37 and earlier grey models
carried out by Swesty\cite{fdsgrey}.

\section{Conclusions and Future Directions}

The issues and work described in this paper fall
far short of offering complete coverage of the
active issues that remain in the area of the
explosion mechanism of core collapse supernovae.
Indeed, we have ignored many important issues such
as magnetic fields, rotation, and neutrino flavor
mixing.  Clearly there is a large gulf of
unexplored physics incorporated in those subjects.

Our own future efforts will be focused in two
areas in the near future.  The first of these is
understanding the effects of numerous microphysics
enchancements that will be added to the models.
The second of these efforts involves extending the
models to 3-D where a more realistic convective
flow structure can arise.

\section*{Acknowledgments}
The authors would like to thank Ed Bachta and Polly Baker of Indiana
University at Indianapolis for their collaborative efforts in
developing the visualization technology that went into Figures
\ref{fig:fig3} and \ref{fig:fig4}.  We gratefully acknowledge the
support of the U.S.\ Dept.\ of Energy, through SciDAC Award
DE-FC02-01ER41185, by which this work was funded.  We are also
grateful to the National Energy Research Scientific Computing Center
(NERSC) for computational support.  Finally, we would like to thank
the National Institute for Nuclear Theory at the University of
Washington for its hospitality in hosting the workshop ``Open Issues
in Core-Collapse Supernovae,'' at which this work was presented in
June 2004.

\appendix


\begin{thebibliography}{0}
\bibitem{BW82} R.\ L.\ Bowers and J.\ R.\ Wilson, {\it Astrophys.\ J.\ Supp.}, 
{\bf 50}, 115 (1982).
\bibitem{bruenn85} S.\ W.\ Bruenn, {\it Astrophys.\ J.\ Supp.}, 
{\bf 58}, 771 (1985).
\bibitem{bm94} S.\ W.\ Bruenn and A.\ Mezzacappa,
{\it Astrophys.\ J.}, {\bf 433}, L45 (1995).
\bibitem{bmd95} S.\ W.\ Bruenn, A.\ Mezzacappa and T.\ Dineva,
{\it Phys.\ Rep.}, {\bf 69}, 256 (1995).
\bibitem{bd96} S.\ W.\ Bruenn and T.\ Dineva,
{\it Phys.\ Rep.}, {\bf 458}, L71 (1996).
\bibitem{bml} A.\ Burrows, Mazurek, T.\ J.\ and J.\ M.\ Lattimer, 
{\it Astrophys.\ J.}, {\bf 251}, 325 (1981).
\bibitem{cvdhb} J.\ Cooperstein, L.\ J.\ van den Horn and E.\ Baron,
{\it Astrophys.\ J.},  {\bf 309}, 653 (1986).
\bibitem{freed} D.\ Z.\ Freedman, {\it Phys.\ Rev.\ D.}, 
{\bf 9}, 1389 (1974).
\bibitem{fw} C.\ L.\ Fryer and M.\ S.\ Warren, {\it Astrophys.\ J.}, 
{\bf 601}, 391 (2004).
\bibitem{engine} M.\ Herant, W.\ Benz, R.\ Hix, C.\ L.\ Fryer and S.\
A.\ Colgate, {\it Astrophys.\ J.}, {\bf 435}, 339 (1994).
\bibitem{hix} R.\ Hix, personal communication, 2004. 
\bibitem{janka03} H.-T.\ Janka, R.\ Buras, K.\ Kifonidis, T.\ Plewa and
M.\ Rampp, in {\it From Twilight to Highlight: The Physics of Supernovae}, 
eds. W.\ Hillebrandt and B.\ Leibundgut, Springer, Berlin, 2003.
\bibitem{keil96} W. Keil, H.-T. Janka, and E. Muller,
{\it Astrophys.\ J.}, {\bf 473}, L111 (1996).
\bibitem{LLPR} Lattimer, J.\ M., C.\ J.\ Pethick, D.\ G.\ Ravenhall, and
D.\ Q.\ Lamb, {\it Nucl.~Phys.~A}, {\bf 432}, 646 (1985).
\bibitem{LS} Lattimer, J.\ M. and F.\ D.\ Swesty, {\it Nucl.\ Phys.\ A}, 
{\bf 535}, 331 (1991).
\bibitem{LP} C.\ D.\ Levermore and G.\ C.\ Pomraning, {\it Astrophys.\ J.}, 
{\bf 248}, 321 (1981).
\bibitem{livne} E.\ Livne, A.\ Burrows, R.\ Walder, I.\ Lichtenstadt
and T.\ Thompson, {\it Astrophys.\ J.}, {\bf 609}, 277 (2004).
\bibitem{mb93a} A.\ Mezzacappa and S.\ W.\ Bruenn,
{\it Astrophys.\ J.}, {\bf 405}, 669 (1993).
\bibitem{mb93b} A.\ Mezzacappa and S.\ W.\ Bruenn,
{\it Astrophys.\ J.}, {\bf 405}, 685 (1993).
\bibitem{mezz98} A.\ Mezzacappa, A.\ C.\ Calder, S.\ W.\ Bruenn, 
J.\ M.\ Blondin, M.\ W.\ Guidry, M.\ R.\ Strayer and A.\ S.\ Umar,
{\it Astrophys.\ J.}, {\bf 493}, 848 (1998).
\bibitem{MBHLSV} E.\ S.\ Myra, S.\ A. Bludman, Y.\ Hoffman, I.\
Lichtenstadt, N.\ Sack and K.\ A.\ Van Riper, {\it Astrophys.\ J.}, 
{\bf 318}, 744 (1987).
\bibitem{RJ} M.\ Ramp and H.-T.\ Janka, {\it Astron.\ \&.\ Astrophys.}, 
{\bf 396}, 361 (2002).
\bibitem{sn1} J.\ M.\ Stone and M.\ L.\ Norman, {\it Astrophys.\ J.\ Supp.}, 
{\bf 80}, 753 (1992).
\bibitem{sn2} J.\ M.\ Stone and M.\ L.\ Norman, {\it Astrophys.\ J.\ Supp.}, 
{\bf 80}, 791 (1992).
\bibitem{sn3} J.\ M.\ Stone, D.\ Mihalas and M.\ L.\ Norman, 
{\it Astrophys.\ J.\ Supp.}, {\bf 80}, 819 (1992).
\bibitem{sb} P.\ Strack and A.\ Burrows, {\it Phys.\ Rev.\ D.}, 
{\bf 71}, 093004 (2005).
\bibitem{slm93} F.\ D.\ Swesty, J.\ M.\ Lattimer and E.\ S.\ Myra,
{\it Astrophys.\ J.}, {\bf 425}, 195 (1994).
\bibitem{FDSTCT} F.\ D.\ Swesty, {\it J.\ Comp.\ Phys.}, {\bf 127}, 118
(1996).
\bibitem{fdsgrey} F.\ D.\ Swesty, in {\it Stellar Evolution, Stellar
Explosions and Galactic Chemical Evolution}, ed. A.\ Mezzacappa,
Institute of Physics Publishing, Bristol, p.\ 539, (1998).
\bibitem{SSS} F.\ D.\ Swesty, D.\ C.\ Smolarski and P.\ E.\ Saylor, 
{\it Astrophys.\ J.\ Supp.}, {\bf 153}, 369 (2004).
\bibitem{wm93} J.\ R.\ Wilson and R.\ W.\ Mayle,
{\it Phys.\ Rep.}, {\bf 227}, 97 (1983).
\bibitem{wm88} J.\ R.\ Wilson and R.\ W.\ Mayle,
{\it Phys.\ Rep.}, {\bf 163}, 63 (1988).
\bibitem{ww} S.\ E.\ Woosley and T.\ A. Weaver, {\it Astrophys.\ J.\ Supp.},
{\bf 101}, 181 (1995).
\bibitem{yb76} W.\ R.\ Yueh and J.\ R. Buchler, 
{\it Astrophys.\ \& \ Space Sci.}, {\bf 39}, 429 (1976).


\end{thebibliography}
\end{document}